\documentstyle[12pt,psfig,epsf]{article}
\newcommand{\be}{\begin{equation}}
\newcommand{\ee}{\end{equation}}
\newcommand{\pr}{\prime}

\newcommand{\f}{\frac}

\newcommand{\bt}{\beta}

\newcommand{\z}{\zeta}

\begin {document}
\thispagestyle{empty}

\hfill NTUA-93/00

\begin {center}
{\Large {\bf Decoupling of Layers in the Three-dimensional 
Abelian Higgs Model}} 
\end{center}

\vspace{3cm}

\begin {center}

{\bf P.Dimopoulos, K.Farakos and G.Koutsoumbas}

\vspace{3cm}

Physics Department

National Technical University, Athens

Zografou Campus, 157 80 Athens, GREECE

\end{center}

\vspace{2cm}

\begin {center}
{\bf ABSTRACT}
\end{center}

The Abelian Higgs model with anisotropic couplings 
in $2+1$ dimensions is studied 
in both the compact and non--compact formulations. 
Decoupling of the space--like planes takes place in the extreme 
anisotropic limit, so charged particles and gauge fields are presumably 
localized within these planes.
The behaviour of the model under the influence of an external magnetic 
field is examined in the compact case and yields further characterization
of the phases.

\vfill

\newpage

\section{Introduction}

The existence of a layered phase in gauge theories with anisotropic 
coupling has been first conjectured by Fu and Nielsen (\cite{Fu}, 
see also \cite{Huang}) in
the early eighties. If the couplings are equal in all directions, 
the Wilson loops are governed by the area law (if the couplings are strong)
or by the perimeter law (if they are weak). If now the couplings are weak
in d dimensions and strong in the remaining one dimension, then the 
model will exhibit Coulomb forces in the d-dimensional subspace and will
confine in the last direction; this means that charged particles 
will be localized within the d--dimensional subspace. 
In the extreme case, where the coupling in the
extra dimension is infinite, charged particles from each subspace will
remain strictly confined to the d--dimensional space--time; in other words, the 
d-dimensional hyperplanes(layers) will decouple from each other and this 
is at the origin of the characterization of this phase as ``layered".

Monte Carlo evidence for a layered structure in pure
U(1) has been given later on \cite{Berman, Altes}. There exists a
lowest dimension for which the layered phase exists: it is the dimension
above which the system has two phases. For the abelian gauge theory
it is at five dimensions that one may first discriminate between the 
Confinement and the Coulomb phases; for the non-abelian case the 
critical dimension is six.

The initial motivation for the study of a model with anisotropic 
couplings has been dimensional reduction with cosmological implications:
if someone lives in one of the d-dimensional hyperplanes of a system in the
layered phase, he has no way of communicating with the remaining layers and
a kind of dimensional reduction is achieved. These considerations have also 
been given a modern version in the context of membrane theories 
and confining (or localization) of
particles within the branes \cite{Randall}.  
On the other hand, the construction 
of the chiral layer is an important ingredient of the 
models of chiral fermions on the lattice \cite{Kaplan}.

In this paper we try to perform an exploratory study of a richer 
model. We treat a (2+1)-dimensional Abelian Higgs model. This has the 
advantage that it is a low-dimensional model with a phase transition (at
large gauge couplings), so it may have a layered phase. In addition, this 
system has a direct relation to the ceramics exhibiting high--$T_{c}$ 
superconductivity \cite{Dorey}. These materials have a layered structure, 
so they are
effectively planar and high--$T_{c}$ superconductivity is considered as a 
two dimensional phenomenon. In this sense the $(2+1)$ -- dimensional 
theory acquires physical relevance. The isotropic version of the model 
has been extensively studied recently \cite{Dimo,Kajantie}, in connection with 
cosmological considerations and Condensed Matter Physics, in which external 
magnetic fields play a predominant role, such as penetration and vortex 
formation.

Another feature of the various phase transitions associated with the
model is that they may be strengthened in the presence of an external 
magnetic field; in addition, the phases of the model with compact action  
may be characterized by the behaviour of the system under the influence 
of the magnetic field. 

Both the compact and non--compact
versions of the model have been examined in this paper and they 
both have characteristics that one can attribute to layer 
decoupling.

\section{Formulation of the model}

The model under study is the Abelian Higgs model in the three-dimensional 
space. Direction $\hat 3,$ corresponding to $z,$ will be singled 
out by couplings that will differ from the rest; later on an external
magnetic field will also point in this direction. 

We first discuss the non-compact version of the model; we proceed with writing
down its lattice action.
$$S=\f{1}{2} \beta _{gs}\sum _{x}F_{12}^2(x)
   +\f{1}{2} \beta _{gt}\sum _{x} [F_{13}^{2}(x)+F_{23}^{2}(x)]
$$
$$
+\f{1}{2} \beta _{hs} \left( \sum _{x}[2 \varphi^*(x)\varphi (x) - \varphi^*(x)U_{\hat 1}(x)\varphi (x+\hat 1) - \varphi^*(x)U_{\hat 2}(x)\varphi (x+\hat 2)]+h.c. \right)
$$
$$
+\f{1}{2} \beta _{ht} \left( \sum _{x}[\varphi^*(x)\varphi (x) - \varphi^*(x)U_{\hat 3}(x)\varphi (x+\hat 3)]+h.c. \right)
$$
\be
+\sum _{x}[(1-2\beta _{R}-2 \beta _{hs}-\beta_{ht})\varphi ^*(x)\varphi (x)
+\beta_{R}(\varphi ^*(x)\varphi (x))^{2}] 
\label{action}
\ee
where $F_{ij} \equiv A_{j}(x+\hat i)-A_{j}(x)-A_{i}(x+ \hat j)+A_{j}(x).$

We have allowed for different couplings in the various directions: the ones
pertaining to directions $\hat 1,~\hat 2$ are given the subscript ``s" for
``space", while direction $\hat 3$ gets couplings with the subscript ``t" for
``time". We stress that the third direction is not really ``time"; it is just 
the direction singled out as explained previously. However, we use this 
notation for convenience; in particular, we refer below to ``time-like" 
plaquettes, links or couplings with this understanding.

For future use we mention here some notations. The link variables 
$U_{\hat k}(x)$ 
are defined as $e^{i g_s \alpha_s A_s}$ or $e^{i g_t \alpha_t A_t}$ 
respectively. They are also written in the form 
$U_{\hat k}(x) = e^{i \theta_{\hat k}(x)},$ since they are complex phases. 
In addition, the scalar fields are also written in the polar form 
$\varphi(x) = R(x) e^{i \chi(x)}.$ The order parameters that we will use are 
the following:
\be
{\rm Space-like~Plaquette:~~~} P_s \equiv <\f{1}{N^3} \sum_x F_{12}^2(x)>
\ee
\be
{\rm Time-like~Plaquette:~~~} P_t \equiv <\f{1}{2 N^3} \sum_x (F_{13}^2(x)
+F_{23}^2(x))>
\ee
\be
{\rm Space-like~Link:~~~} L_s \equiv <\f{1}{2 N^3} \sum_x 
[\cos(\chi(x+\hat 1) +\theta_{\hat 1}(x)-\chi(x))
\ee
\be
+\cos(\chi(x+\hat 2) +\theta_{\hat 2}(x)-\chi(x))]>
\ee
\be
{\rm Time-like~Link:~~~} L_t \equiv <\f{1}{N^3} \sum_x 
\cos(\chi(x+\hat 3) +\theta_{\hat 3}(x)-\chi(x))>
\ee
\be
{\rm Higgs~field~measure~squared:~~~} \rho^2 \equiv \f{1}{N^3} \sum_x R^2(x)
\ee

We now proceed to the na\"ive continuum limit (at tree level)
of the lattice action [10]. The first step is to rewrite the action in terms 
of the continuum fields (denoted by a bar):

$$
\varphi = {\overline \varphi} (\f{2 a_s}{\beta_{hs}})^{1/2}, 
$$
$$
~~ A_1 = a_s {\overline A_1}, ~~A_2 = a_s {\overline A_2}, 
~~A_3 = a_t {\overline A_3}.
$$
This means that the time-like part of the pure gauge action is rewritten in the 
form: $\beta_{gt}a_{t} \sum a_{s}^{2}a_{t}[({\overline F}_{13})^{2}+ 
({\overline F}_{23})^{2}] \rightarrow \beta_{gt}a_{t} \int d^{3}x 
[({\overline F}_{13})^{2}+ 
({\overline F}_{23})^{2}]$. On the other hand the space--like part is:
$\beta_{gs}\frac{a_{s}^{2}}{a_{t}} \sum a_{s}^{2}a_{t}({\overline F_{12}})^{2} 
\rightarrow \beta_{gs} \frac{a_{s}^{2}}{a_{t}} \int d^{3} x 
({\overline F_{12}})^{2}$. If we define
\be
\beta_{gs} \equiv \frac{a_{t}}{g_{s}^{2}a_{s}^{2}},
~~\beta_{gt} \equiv \frac{1}{g_{t}^{2} a_{t}},
\ee
the resulting continuum action reads:
$$\f{1}{2} \int d^3 x {\left [ \frac{1}{g_{s}^{2}}({\overline F_{12}})^{2}+\frac{1}{g_{t}^{2}}[({\overline F_{13}})^{2}+({\overline F_{23}})^{2}] \right ]}$$
Defining $\gamma_{g} \equiv (\frac{\beta_{gt}}{\beta_{gs}})^{1/2}$ and using
the definitions of $\beta_{gs}$, $\beta_{gt}$ we find that $\gamma_{g}=
\frac{g_s}{g_t}\frac{a_s}{a_t}$. We denote by $\xi$ the important ratio 
$\frac{a_s}{a_t}$ of the two lattice spacings (the correlation anisotropy 
parameter) and finally derive the relation:
$$\gamma_{g}=\left( \f{\bt_{gt}}{\bt_{gs}} \right)^{(1/2)} 
= \frac{g_s}{g_t} \xi.$$
Along the same lines, one may rewrite the scalar sector of the action in the 
form:
$$\int d^3 x [|D_1 {\overline \varphi}|^2 
+|D_2 {\overline \varphi}|^2
+\f{\gamma_\varphi^2}{\xi^2} |D_3 {\overline \varphi}|^2+
m^2 {\overline \varphi}^* {\overline \varphi} 
+ \lambda ({\overline \varphi}^* {\overline \varphi})^2 ]
$$

We have used the notations: 
$$ \gamma_\varphi \equiv \left( \f{\beta_{ht}}{\beta_{hs}} \right)^{1/2},$$
$$ m^2 a{_s}^{2} \equiv \frac{2}{\beta_{hs}}(1-2\beta_{R}
-2\beta_{hs}-\beta_{ht}),~~\lambda a_s = \f{4\beta_R}{ \beta_{hs}^2 \xi}.$$

Of course, the dependence of the space-like and time-like couplings on the 
lattice spacings $a_s~{\rm and}~a_t$ is a very interesting 
issue, which has been pursued in the context of QCD \cite{Kar,
Engels, Montvay}. 
Neglecting quantum corrections we have $\gamma_\varphi=\gamma_g=\xi.$
In this paper we would like to explore the
decoupling of the space-like planes, so in principle we should have a 
relation for the variation of $\xi$ as a function of 
$\gamma_\varphi$ and $\gamma_g$, with the quantum corrections 
properly taken into account. We defer this study to a later stage and
for the time being we adopt the tree level scaling refered to previously.
In particular, we will vary the quantities 
$\beta_{gs},~\beta_{gt},~\beta_{hs},~\beta_{ht}$ in such a way that 
we have $$ \gamma_\varphi=\gamma_g \equiv \zeta.$$
We have used the letter $\zeta$ for the common ratio of the couplings, so 
our assumption reads: $\xi=\zeta$ and $g_s=g_t \equiv g.$

The parameter $\beta_R$ is found 
from the equation $\beta_R = \f{x \beta_{hs}^2}{4 \beta_{gs}},$ 
using the fixed value $x=2$ for the parameter
$x \equiv \f{\lambda}{g^2}.$ 

The compact action is defined as:

$$S_{compact}=\beta _{gs}\sum _{x}(1-\cos F_{12}(x))
   +\beta _{gt}\sum _{x} \left [ (1-\cos F_{13}(x))+(1-\cos F_{23}(x)) \right ]
$$
$$
+\beta _{hs}\sum _{x}[2 \varphi^*(x)\varphi (x)
- \varphi^*(x)U_{\hat 1}(x) e^{i A_{\hat 1, ext}(x)} \varphi (x+\hat 1)
- \varphi^*(x)U_{\hat 2}(x) e^{i A_{\hat 2, ext}(x)} \varphi (x+\hat 2)]
$$
$$
+\beta _{ht}\sum _{x}[\varphi^*(x)\varphi (x)
- \varphi^*(x)U_{\hat 3}(x) e^{i A_{\hat 3, ext}(x)} \varphi (x+\hat 3)]
$$
\be
+\sum _{x}[(1-2\beta _{R}-2 \beta _{hs}-\beta_{ht})\varphi ^*(x)\varphi (x)
+\beta_{R}(\varphi ^*(x)\varphi (x))^{2}]
\label{compactaction}
\ee
The difference from the non-compact version lies in the gauge kinetic terms;
however, since we will treat the compact system in an external magnetic field,
we have taken the opportunity to also include in the action the 
background potential $A_{\hat k, ext}(x),$ which will generate the external
field. There are many ways to impose such an external magnetic field
\cite{Degrand,Kajantie,Dam}; for example, one might include 
it only in the interaction term with the matter fields; that is, the matter 
fields interact with both the quantum and the background fields.
This has been our choice, as can be seen from equation (\ref{compactaction}).
 
The order parameters in this case are defined in a very similar way as in 
the non-compact case. We state the expressions which differ:
\be
{\rm Space-like~Plaquette:~~~} P_s \equiv <\f{1}{N^3} \sum_x 
cos F_{12}(x)>
\ee
\be
{\rm Time-like~Plaquette:~~~} P_t \equiv <\f{1}{2 N^3} 
\sum_x [cos F_{13}(x) + cos F_{23}(x)]>
\ee

\section{Algorithms}
 We used the Metropolis algorithm for the updating of both the gauge and
the Higgs field. It is known that the scalar fields have much longer
autocorrelation times than the gauge fields. Thus, special care must be
taken to increase the efficiency of the updating for the Higgs field. We
made the following additions to the Metropolis updating procedure 
\cite{Dimo}:

{\bf a) Global radial update:} We update the radial part of the Higgs field
by multiplying it by the same factor at all sites: $R(x)
\rightarrow e^{\xi }R(x),$ where $\xi \in [-\varepsilon ,
\varepsilon]$ is randomly chosen. The quantity $\varepsilon$ is adjusted
such that the acceptance rate is kept between 0.6 and 0.7. The probability
for the updating is $P(\xi )=$ min$\{1,\exp (2V\xi -\Delta S(\xi)) \}$ where
$\Delta S(\xi )$ is the change in action, while the $2V\xi$ term comes
from the change in the measure.

{\bf b) Higgs field overrelaxation:} We write the Higgs potential at
$x$ in the form: 
\be
V(\varphi (x))=-{\bf a} \cdot {\bf F} +
R^{2}(x)+\beta _{R}(R^{2}(x)-1)^{2} 
\ee
where
$${\bf a} \equiv \left( \begin{array}{c} R(x) \cos \chi (x)\\
                                   R(x) \sin \chi (x)
                   \end{array}
                               \right),$$
$$
{\bf F} \equiv \left(\begin{array}{c} F_1\\F_2 \end{array}
\right),$$
$$ F_1 \equiv \beta_{hs} \left[ \sum_{i=1,2} [R(x+ \hat i)\cos(\chi (x+\hat i)+\theta_{\hat i}(x)) \right] 
+\beta_{ht} R(x+ \hat 3)\cos
(\chi (x+\hat 3)+\theta_{\hat 3}(x)), 
$$
$$
F_2 \equiv \beta_{hs} \left[ \sum_{i=1,2} R(x+\hat i)\sin (\chi (x+\hat i)+\theta_{\hat i}(x)) \right] 
+\beta_{ht} R(x+\hat 3)\sin (\chi (x+\hat 3)+\theta_{\hat 3}(x)).
$$
We can perform the change of variables: $({\bf a},{\bf F}) \rightarrow
(X,F,{\bf Y})$ ,where
\be
F \equiv |{\bf F}|,~~~ {\bf f} \equiv \frac{{\bf F}}{\sqrt{F_1^2 + F_2^2}},~~~
X \equiv {\bf a} \cdot {\bf f},~~~{\bf Y} \equiv {\bf a} - X {\bf f}.
\ee

The potential may be rewritten in terms of the new variables:
\be
\bar V(X,F, {\bf Y})=-XF +(1+2\beta _{R}({\bf Y}^{2}-1)) X^{2}
+{\bf Y}^{2}(1-2\beta_{R})+\beta _{R}(X^{4}+{\bf Y}^{4}).
\ee
The updating of ${\bf Y}$ is done simply by the reflection:
\be
{\bf Y} \rightarrow {\bf Y}'= -{\bf Y}.
\ee

The updating of X is performed by solving the equation:
\be
\bar V(X^\pr,F, {\bf Y})= \bar V(X,F, {\bf Y})
\ee
with respect to $X^\pr.$
Noting that $X^\pr=X$ is obviously a solution, we may factor out the 
quantity $X^\pr - X$ and reduce the quartic equation into a cubic one, which
may be solved.
The change $X \rightarrow X'$ is accepted with probability:
 $P(X')=$ min$\{P_0,1\},$ where $P_0 \equiv
 \frac{\partial \bar V(X,F,{\bf Y})}{\partial X}/\frac{\partial \bar
 V(X',F',{\bf Y'})}{\partial X'}$.

\section{Results}

\subsection{Runs with fixed $\zeta$}

\begin{figure}
\centerline{\hbox{\psfig{figure=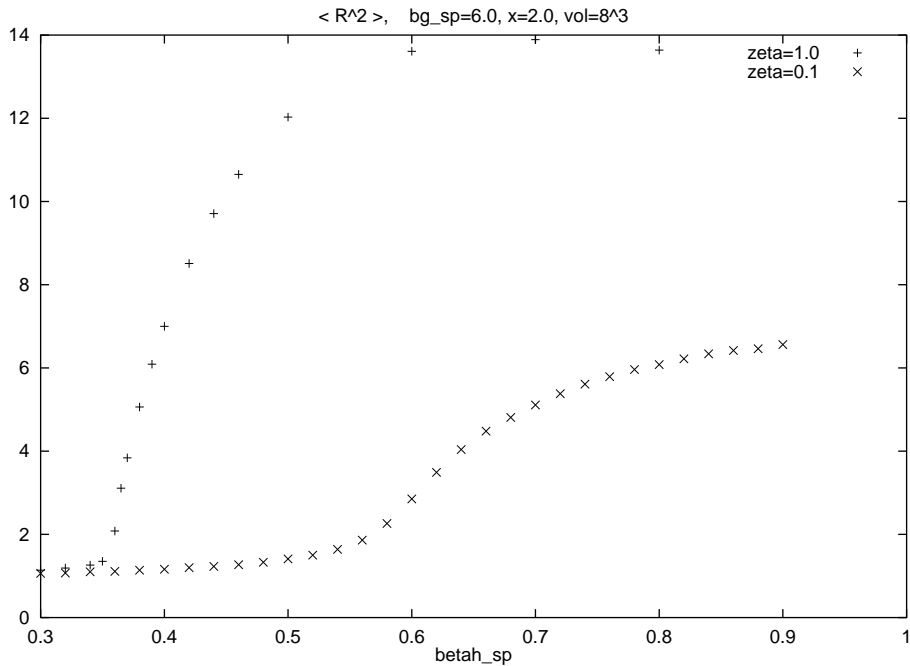,height=9cm,angle=-90}}}
\caption[f1]{$\rho^2$ versus $\beta_{hs}$ for the S model (
$\zeta=1.0$) and the A model ($\zeta=0.1$) for a $8^3$ lattice with
$\beta_{gs}=6.0$ and $x=2$.}
\label{f1}
\end{figure}

In this set of measurements we set $\beta_{gs}=6.0~~$,$x=2$,
and let $\beta_{hs}$ run. The remaining coupling 
constants vary according to the value of $\z,$ that is:
\be
\beta_{ht}= \zeta^{2} \beta_{hs},~~\beta_{gt}= \zeta^{2} \beta_{gs},~~
\beta_{R}=\frac{x \beta_{hs}^{2}}{4 \beta_{gs}}.  
\ee
We recall that $\zeta$ is the ratio: $\zeta=\sqrt {\frac{\beta_{gt}}{\beta_{gs}}}=
\sqrt{\frac{\beta_{ht}}{\beta_{hs}}}$. 
In figure \ref{f1} we show a comparison of the behaviour of the 
measure of the scalar field $\rho^2 $  
for the symmetric $(\z=1.0)$ and a highly asymmetric model $(\z=0.1)$, that
we call ``S(ymmetric) model" and ``A(symmetric) model" respectively.
We have used the compact action in both models. The S model ($\zeta=1.0$)
shows that the system undergoes a phase transition starting at $\beta_{hs} \simeq
0.35$ from the 3D Coulomb phase to the 3D Higgs phase. (We note here that $\rho^2$
 is not going to infinity for big $\beta_{hs}$, which is the case in several
treatments; the reason is that $\beta_{R}$ is not fixed, but increases 
with increasing $\beta_{hs}$). Then there is a transition region from 
$\beta_{hs} \simeq 0.35$ to $\beta_{hs} \simeq 0.50$; for $\beta_{hs} \geq 0.50$
the system is in the Higgs phase, characterized by a large value of $\rho^2$. 

The corresponding transition for the A model ($\zeta=0.1$) moves towards bigger 
values of $\beta_{hs}$ and it seems smoother. 
In the latter case the ``time--like" 
coupling constants are $\zeta^2=0.01$ times smaller than their ``space--like" 
partners. Let us note that for $\zeta=0$ there will be no communication at all 
among the planes. 
This lends support to the assumption that for $\zeta=0.1$, that is
$\zeta^2=0.01$, we may have reached the limit of decoupled layers; indeed,
we have checked that for an even smaller ratio of couplings ($\zeta=0.01$) the 
curve for the A model does not change much. This presumably means that 
the time--like separation $a_{t}$ of the spatial planes is already much bigger 
than the spatial lattice spacing $a_{s}$. This is further supported by the 
measurement of other quantities, which depend on the direction (see, for 
example, figure 2 and especially figure 3 below). It turns out that 
the quantities
$P_{s}$, $L_{s}$ are much bigger than $P_{t}$, $L_{t}$ for all the range of
$\beta_{hs}$, indicating that the quantities  related to the
communication of the planes are negligible in this case, as compared against the similar 
quantities within the layers. It seems safe to assume that the region 
$\beta_{hs} \leq 0.55$ corresponds to a Coulomb phase, where the layers 
are decoupled: this is equivalent to the Coulomb phase of the corresponding 
{\em two--dimensional} model. The transition region extends from $\beta_{hs} \simeq 
0.55$ to $\beta_{hs} \simeq 0.65$. and then we have a Higgs phase for the A model:
 it has smaller $\rho^2$ than the one of the S model and the quantities related
to the third dimension are very small. This presumably means that the layers 
are decoupled also in this phase, so the picture is that we have moved effectively to the Higgs 
phase of the corresponding {\em two--dimensional} model. Thus we may say 
in brief that in the A model we see a transition from {\em the 2D Coulomb
phase to the 2D Higgs phase}. 
Let us stress here that we have been using the words Higgs ``phase" and 
``phase" transition in a rather loose meaning when refering to the
asymmetric model. We have not checked 
that the passage from one region to the other is actually a
phase transition: it may be a crossover, so the exact characterization
is open for the time being. We should remark that the critical parameter 
for the Higgs phase transition is of order $\frac{1}{d}$, where d is the 
space--time dimension. Thus, it is not accidental that the isotropic model
has a phase transition at $\beta_{hs} \simeq 0.35$; this is close to the expected 
value, since d=3. On the other hand, the anisptropic model is effectively 
two--dimensional, so one should expect a value approximately equal to
$\frac{1}{2}$, in semi--quantitative agreement with our results.

We note that the relative position of the transitions means that for 
$\beta_{hs} \le 0.35$ the system lies in the Coulomb phase for either
value of $\z;$ for $ \beta_{hs}=0.50$ the S model lies in its 3D Higgs phase,
while the A model is still in its 2D Coulomb region; for $\beta_{hs}=0.70$ both
models lie in their respective Higgs phases. The region between the two curves 
of figure 1 is presumably full of other curves corresponding to all values
of $\zeta$ between 0.1 and 1. Later on we will consider the behaviour of several
quantities with fixed $\beta_{hs}$ but varying $\zeta$; figure 1 suggests that 
we may use $\beta_{hs}=0.5$ and $\beta_{hs}=0.7$, based on the comments 
we made above.

Figure 2 shows the same models as above, but the space-like link is depicted
rather than $\rho^2;$ in addition, we have shown the results for both the
compact and the non-compact version of the model. The changes take place 
at the same values of $\beta_{hs}$, as in figure 1. 
The results of the non-compact model are very close to the ones 
of the compact model, the most sizable differences being observed 
in model S, where the discrepancy reaches the value 0.2 in the region of 
the phase transition. In model A the differences are really small.

\begin{figure}
\centerline{\hbox{\psfig{figure=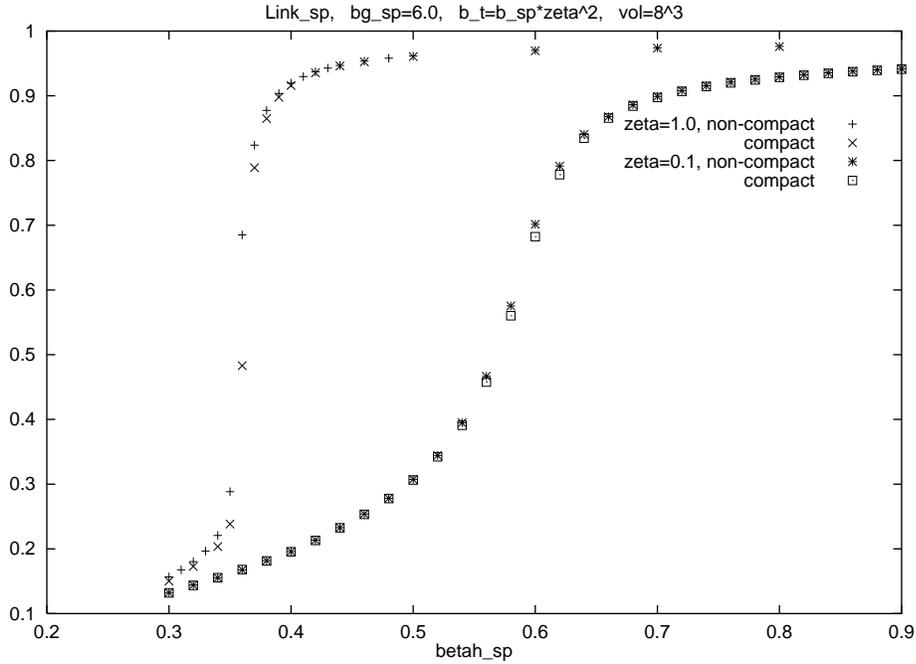,height=9cm,angle=-90}}}
\caption[f2]{$L_{s}$ versus $\beta_{hs}$ for the S model ($\zeta=1.0$)
and the A model ($\zeta=0.1$) for a $8^3$ lattice with $\beta_{gs}=6.0$ 
and $x=2.0$.}
\label{f2}
\end{figure}

In figure 3 we display the time-like link and plaquette only for the 
{\em compact} version of model A $(\z=0.1),$ to get a more concrete idea 
of the layered phase. The parameters vary in the same way as above. 
We find that the time-like link starts increasing around $\beta_{hs}=0.6,$
which is in the region where the passage in model A is located in
figures 1 and 2. The time-like plaquette $P_{t}$ does not seem to ``feel" the
passage and assumes a constant value. But the most important 
characteristic is that both quantities are really very small:
their maximal value is $0.03,$ while their space-like partners $L_s$ and $P_s$ 
take on much larger values: $0.13<L_s<1.0,~0.90<P_s<1.0.$ This is 
of course a consequence of our choice of bare parameters and confirms the 
fact that quantities related to communication between the planes are
almost negligible in this region of the parameter space.

\begin{figure}
\centerline{\hbox{\psfig{figure=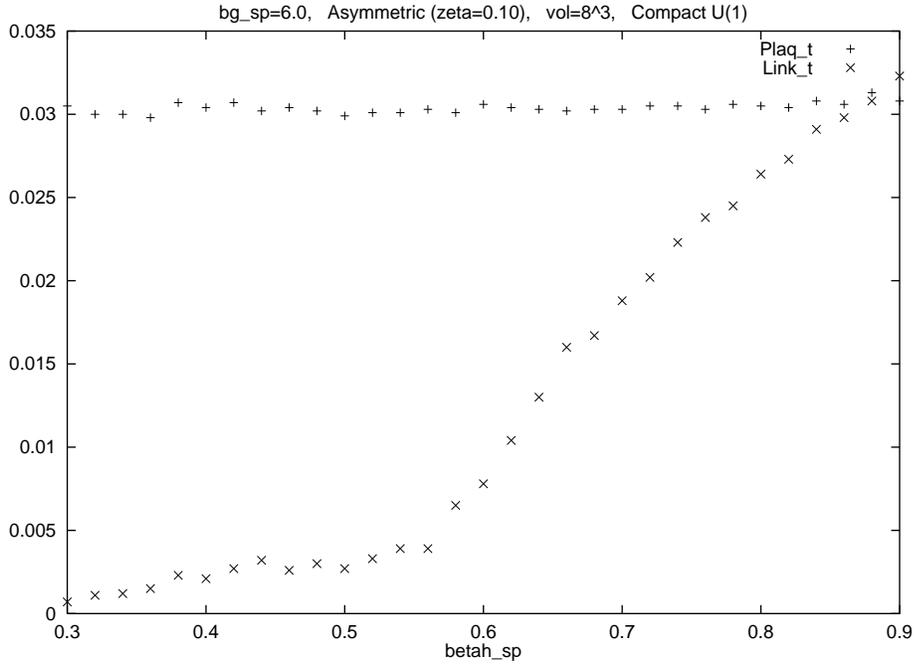,height=9cm,angle=-90}}}
\caption[f3]{Time--like link and plaquette versus $\beta_{hs}$ for
the A model ($\zeta=0.1$) for a $8^3$ lattice with $\beta_{gs}=6.0$ 
and $x=2.0$.}
\label{f3}
\end{figure}

\newpage

It is possible to show in a very explicit way the decoupling of the planes
if we restrict ourselves to the Coulomb phase. The Wilson loops $W(N,N)$ can
be calculated analytically in the 2D (pure) U(1) model and the resulting 
values can then be compared to the (space-like) plaquettes found by 
Monte Carlo simulations of the asymmetric Higgs model  
in the extremely layered Coulomb phase. 
In the table that follows we have put the results for the Wilson loops 
with sizes $N \times N$ from $N=1$ to $N=6.$ The values in parentheses 
are the statistical errors in the last one or two digits.
The values of the parameters are reported in the table and the quantity 
$\z$ is set to 0.1. We also present in the table the values $W(N,N)$ from the
analytical expressions \cite{Kogut}:
$W(N,N) = (\frac{I_1(\beta_{gs})}{I_0(\beta_{gs})})^{(N \times N)}.$ 
It is obvious
that the system has moved to essentially two-dimensional dynamics, since
most of the discrepancies lie within the error bars of the Monte Carlo data.

\[
\begin{array}{||c|c|c||}
\hline
\multicolumn{3}{||c||}{{\rm Table~~1}} \\ 
\hline
\multicolumn{3}{||c||}{{\rm Volume}=12^3,~\z=0.1,~\beta_{hs}=0.30,~x=2.0} \\ 
\hline
{\rm Loop~size}    &{\rm Measured}    &{\rm Analytic~(2D,~\cite{Kogut})}   \\
\hline
\multicolumn{3}{||c||}{\beta_{gs}=2.0} \\
\hline
 1 \times 1           &0.6981(1)    &0.697775     \\
 2 \times 2           &0.2380(2)    &0.237061     \\
 3 \times 3           &0.0395(2)    &0.0392136    \\
 4 \times 4           &0.0033(2)    &0.00315823   \\
 5 \times 5           &0.0002(2)    &0.000123845  \\
 6 \times 6           &0.0001(1)    &0.0000023645 \\
\hline
\multicolumn{3}{||c||}{\beta_{gs}=4.0} \\
\hline
 1 \times 1           &0.8635(1)    &0.863523     \\
 2 \times 2           &0.5562(2)    &0.556026     \\
 3 \times 3           &0.2669(4)    &0.266971     \\
 4 \times 4           &0.0955(5)    &0.0955827   \\
 5 \times 5           &0.0251(4)    &0.0255178  \\
 6 \times 6           &0.0053(2)    &0.00507988 \\
\hline
\multicolumn{3}{||c||}{\beta_{gs}=6.0} \\
\hline
 1 \times 1           &0.9123(1)    &0.912359     \\
 2 \times 2           &0.6931(2)    &0.692889     \\
 3 \times 3           &0.4384(4)    &0.438019    \\
 4 \times 4           &0.2307(9)    &0.230491   \\
 5 \times 5           &0.1008(13)    &0.10096  \\
 6 \times 6           &0.0360(11)    &0.0368106 \\
\hline
\multicolumn{3}{||c||}{\beta_{gs}=8.0} \\
\hline
 1 \times 1           &0.9352(1)    &0.935235     \\
 2 \times 2           &0.7655(1)    &0.76504     \\
 3 \times 3           &0.5490(3)    &0.54738    \\
 4 \times 4           &0.3454(6)    &0.342559   \\
 5 \times 5           &0.1909(9)    &0.18751  \\
 6 \times 6           &0.0925(10)   &0.089775 \\
\hline
\end{array}
\]

\subsection{Runs with varying $\z$}

Now we come to a somewhat more detailed examination of the phase diagram
where the ``space-like" coupling constants are kept fixed and the ratio $\z$ 
changes, also forcing $\beta_{gt}$ and $\bt_{ht}$ to change. Of course,  
the result will strongly depend on the region in which the fixed ``space--like"
coupling constants lie; we will present results for the two interesting 
choices that we described in the discussion of figure 1: $\beta_{gs}=6.0$ 
and the values 0.5 and 0.7 for $\beta_{hs}.$ 

\begin{figure}

\centerline{\hbox{\psfig{figure=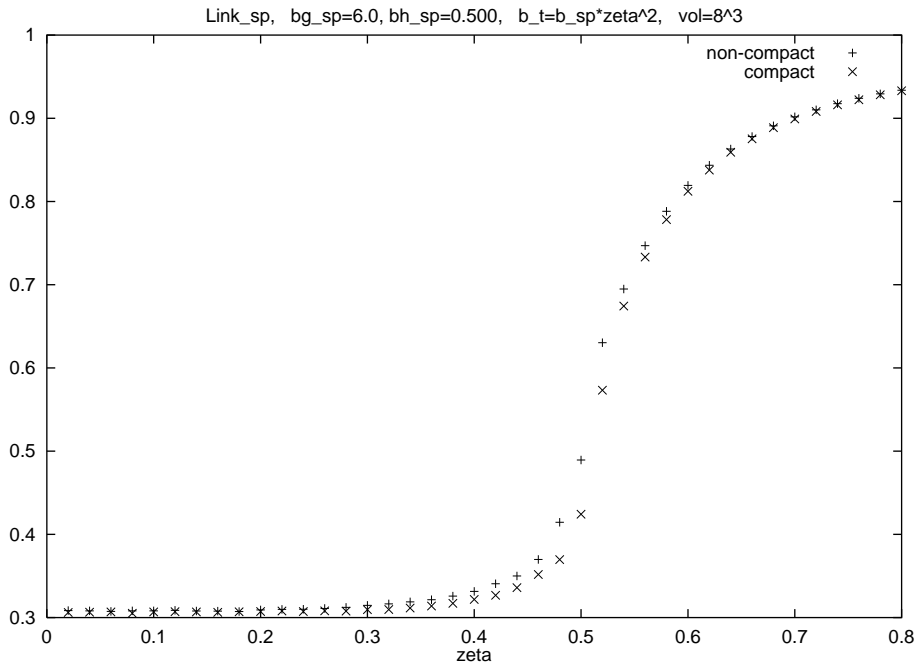,height=9cm,angle=-90}}}
\caption[f4]{$L_{s}$ versus $\zeta$ for $\bt_{gs}=6.0$ and 
$\beta_{hs}=0.500$.}
\label{f4}
\end{figure}

In figure \ref{f4} 
we show the space-like link $L_s$ for the case with $\beta_{hs} = 0.5;$ 
again both compact and non-compact results are displayed. 
The parameter $\zeta$ starts from zero, where we expect a Coulomb phase with 
fully separated layers, that is a 2D model, to 0.8, 
which corresponds to a model close to the symmetric 3D one. 
The differences between the compact and the non--compact version are
localized in the transition region. The non--compact results are somewhat
bigger, the maximum difference being about 0.05, but qualitatively the results 
are similar for both versions.
The transition from the symmetric to the broken phase takes
place at about $\z \simeq 0.5,$ where 2D Coulomb becomes 3D Higgs:
The behaviour of the space-like link may be interpreted with the help of
figure 2. As we change $\zeta$, we move continuously to new curves, of the
sort of the ones in figure 2, but characterizing the $\z$ under study. 
When $\z=0,$ the corresponding curve (close to the lowest one in figure 2)
will inform us that the value $\beta_{hs} = 0.5$ corresponds to the Coulomb
phase for this value of $\z.$ The curves for larger $\z$ will differ from the
initial one in an important aspect: the transition region will move to the 
left (to smaller $\beta_{hs}$) and will be steeper, until finally they 
will coincide with the uppermost curve in figure 2. When the transition region
reaches $\beta_{hs} = 0.5,$ the value of $L_s$ will increase abruptly and
the system will pass over to a Higgs region. It is not clear at this
point whether this is 2D or 3D Higgs, but this will be clarified by figure 5,
which follows.

The time-like link $L_t,$ shown in figure 5 both for the compact and 
non-compact versions, 
seems a more sensitive detector of the phase transition: it changes 
already in the region $0 < \z < 0.5,$ where the space-like link is essentially
constant, and its variation is bigger, since it starts from zero. Of
course, it has to become equal to the space-like link (about one) 
in the limit $\z \to 1,$ since there one regains the symmetric model.  
The non-compact results lie systematically above the compact ones. 
The comparison
of this figure with the previous one shows that the time--like link
also gets its 3D value at the same time when the space--like link jumps.
This fact indicates that we have a transition from the 2D Coulomb phase
to a 3D Higgs, that is, both changes (in dimension and in the breaking 
of symmetry) appear to happen simultaneously.

\begin{figure}
\centerline{\hbox{\psfig{figure=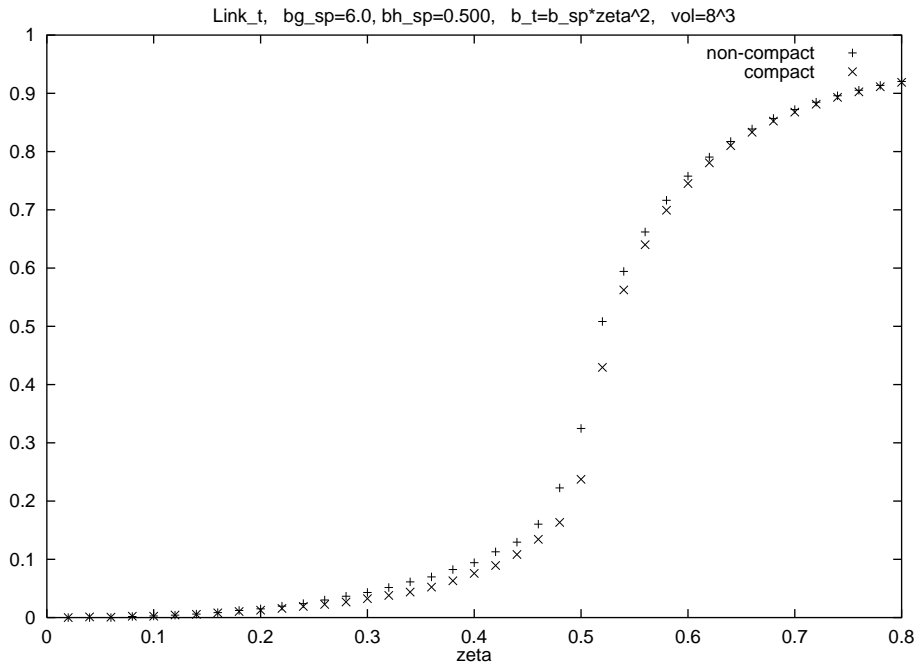,height=9cm,angle=-90}}}
\caption[f5]{$L_{t}$ versus $\zeta$ for $\bt_{gs}=6.0$ and 
$\beta_{hs}=0.500$.}
\label{f5}
\end{figure}

\begin{figure}
\centerline{\hbox{\psfig{figure=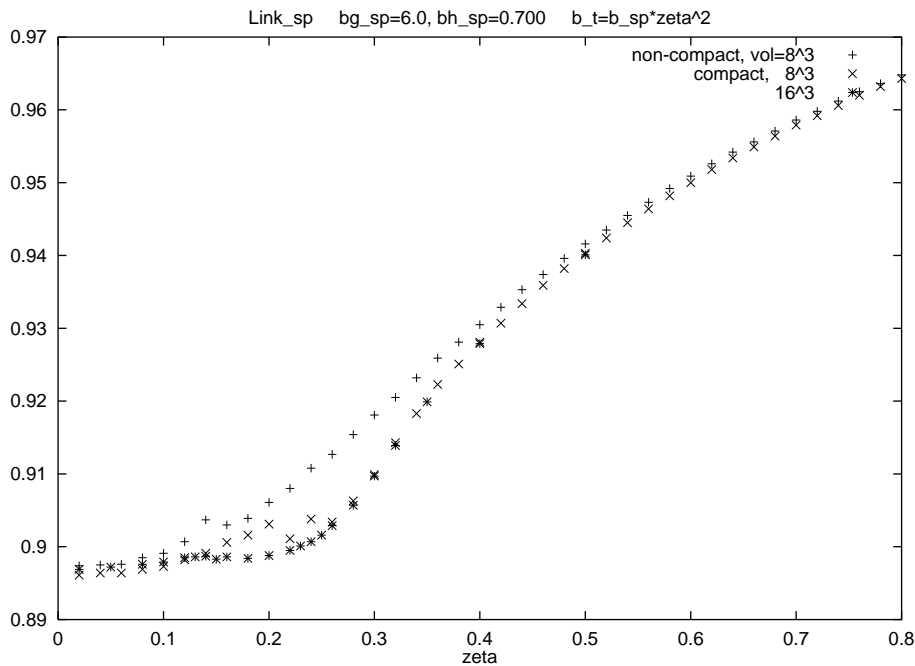,height=9cm,angle=-90}}}
\caption[f6]{Compact and non--compact $L_{s}$ versus $\zeta$ for
$\bt_{gs}=6.0$ and $\beta_{hs}=0.700$.}
\label{f6}
\end{figure}

In figure \ref{f6} we show the space-like link $L_s$ versus $\zeta$ for the 
second of the two chosen values, $\beta_{hs}=0.7.$ 
In this case $\beta_{hs}$ lies to the right of the transition 
already for the curve corresponding to $\z=0,$ as may be seen in figure 2. 
Thus we do not expect a phenomenon as in figure 4, where the 
transition region was passing through $\beta_{hs}.$
We start from a layered Higgs phase already at $\z=0$ and we may
isolate the transition from the 2D to the 3D Higgs phase 
(the space--like link has big values, from 0.90 to 0.97, 
in all the range of $\zeta,$ in striking contrast with figure 4).
The transition happens at $\beta_{hs} \simeq 0.25$ for the compact case.  
It seems smooth, however one may observe a volume dependence in the
results of figure 6, where we have put mesurements from both 
$8^3$ and $16^3$ lattices. In addition, we remark that the points for the compact
$8^3$ have a non--monotonous region between 0.15 and 0.25; this disappears in the
$16^3$ lattice, so we interpret it as finite size artifact. The same effect is seen
in the non--compact $8^3$ case, but seems unimportant, in view of the 
experience with the compact case. The behaviour of the non--compact $8^3$ is
smoother than its compact partner.

Figure 7 contains the results for the time-like link $L_t$ versus $\z$ for
the two volumes $8^3$ and $16^3.$ This quantity is a good indicator of 
the 2D to 3D phase transition, as we said above; its volume 
dependence is almost negligible, but the difference between the 
compact and non--compact version is very pronounced at the region 
of the transition. The non--compact version appears steeper, 
in contrast with the previous figure 6. 

\begin{figure}
\centerline{\hbox{\psfig{figure=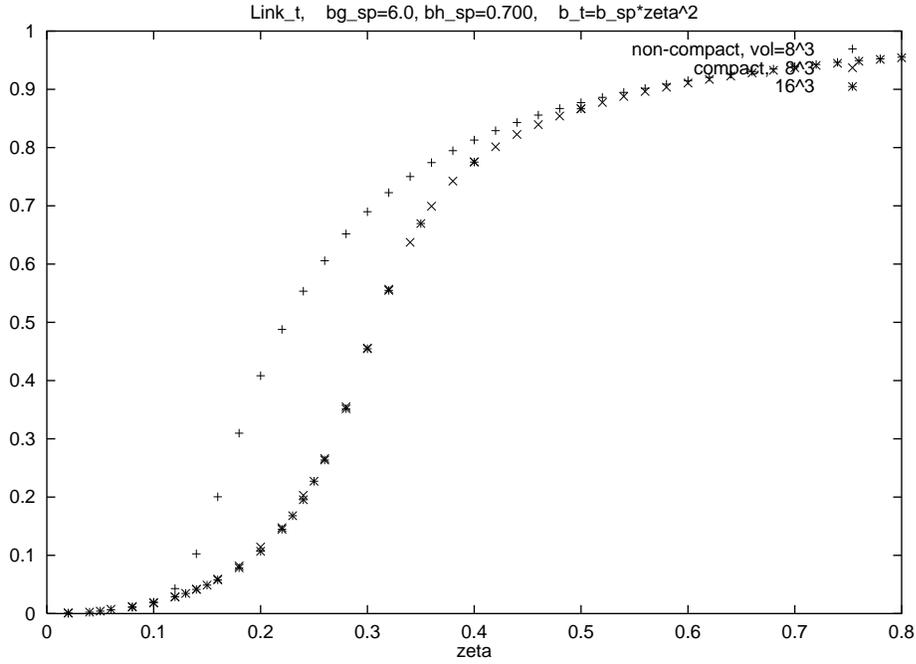,height=9cm,angle=-90}}}
\caption[f7]{Compact and non--compact $L_{t}$ versus $\zeta$ for
$\bt_{gs}=6.0$ and $\beta_{hs}=0.700$.}
\label{f7}
\end{figure}

\subsection{Vortex dynamics in the compact version}
   
We will now consider the system under the influence of a homogeneous 
external magnetic field; thus one should construct first a lattice
version of the homogeneous magnetic field. This has already been 
done before in \cite{Dam} in connection with the abelian Higgs model, 
or (2+1) QED \cite{fkm}.
We follow a slightly different prescription,
which we describe below.

Since we would like to impose an external homogeneous
magnetic field in the $z$ direction, we
choose the external gauge potential in such a way that the plaquettes in
the $xy$ plane equal B, while all other plaquettes equal zero.
One way in which this can be achieved is through the choice:
$A_z(x, y, z)=0$,  for all integers $x, y, z,$ running from 1 to N and
\be
A_x(x, y, z) =-\frac{B}{2} (y-1), x \ne N,~  
A_x(N, y, z)  =-\frac{B}{2} (N+1) (y-1),
\ee
\be
A_y(x, y, z) =+\frac{B}{2} (x-1), y \ne N,~
A_y(x, N, z) =+\frac{B}{2} (N+1) (x-1).
\ee
where $N^3$ is the number of points on the (cubic) lattice and the 
coordinates $x, y, z$ are integers running from 1 to N.  
It is trivial to check out that all plaquettes starting at $(x, y, z),$
with the exception of the one starting at $(N, N, z),$ equal $B.$ The latter 
plaquette equals $(1-N^2) B = B -(N^2 B).$ One may say that the flux
is homogeneous over the entire $xy$ cross section of the lattice and
equals $B.$ The additional flux of $-(N^2 B)$ can be understood by the
fact that the lattice is a torus, that is a closed surface, and the 
Maxwell equation
${\bf \nabla \cdot B} =0$ implies that the magnetic flux through the lattice
should vanish. This means that, if periodic boundary conditions are
used for the gauge field, the total flux of any configuration should be 
zero, so the (positive, say) flux $B,$ penetrating
the majority of the plaquettes, will be accompanied by a compensating
negative flux $-(N^2 B)$ in a single plaquette.
This compensating flux should be ``invisible", that is it should have no
observable physical effects. This is the case if the flux is an integer
multiple of $2 \pi: N^2 B =n 2 \pi \to B=n \frac{2 \pi}{N^2},$ 
where $n$ is an integer. Thus we may say (disregarding the ``invisible" 
flux) that the magnetic field is homogeneous over the entire cross section
of the lattice. The integer $n$ may be chosen to lie in the 
interval $[0, \frac{N^2}{2}],$ with the
understanding that the model with integers $m$ between $\frac{N^2}{2}$ 
and $N^2$ is equivalent to the model with integers 
taking on the values $N^2-n,$ which are among the
ones that have already been considered. It follows that the magnetic field 
strength B in lattice units lies between 0 and $\pi.$ The physical 
magnetic field $B_{phys}$ is
related to $B$ through $B=g_{s} a_s^2 B_{phys},$ and the physical field may
go to infinity letting the lattice spacing $a_{s}$ go to zero,
while $B$ is kept constant.

One of the first questions which may be asked in connection with the 
existence of the external filed is of course its penetration in 
the bulk of the lattice. It is this question which is treated in table 2.
The first column in the table is the integer determining the magnetic field;
according to the previous discussion, its range is from 0 to 32 for the 
$8^3$ lattice that we have used (in table 2 we show the results up to 
$n=20$ only). Apart from $P_{s}$ and $P_{t}$ we show 
the expectation value $P_{s}(B)$ of the space--like plaquette which 
includes the external field. Its definition reads:
$P_{s}(B) \equiv <\frac{1}{N^3} \sum_{x} \cos (F_{12}(x)+B)>$.
The last column contains the number $Q$ of vortices that penetrate into the
lattice to shield the external magnetic field (Meissner effect). 
For this topological number $Q$ we use the definition of \cite{Dam}. 
We consider the space--like plaquette $P_s(x)$ and define the quantity 
$\hat P_s(x),$ and the topological number $q_s(x)$ for the space--like 
plaquette at position $x$ through the equality:
$$ P_s(x)=\hat P_s(x) + 2 \pi q_s(x)$$ and the demand that $\hat P_s$ lies 
in the interval $(-\pi, +\pi].$ The topological number $Q$ is the sum of 
the quantities $q_s(x)$ over space--like plaquettes with fixed z-coordinates: 
$$Q \equiv \sum_{xy} q_s(x).$$

We show the results for a symmetric $(\z=1.0)$ $8^3$ lattice,
with three different values for $\beta_{hs}.$

For $\bt_{hs}=0.200$
(which corresponds to the Coulomb phase) we observe that
neither the space-like nor the time-like plaquettes depend on the
magnetic field. (The two kinds of plaquettes could in principle
be different even for a symmetric lattice, because of the presence
of the external field). The same is true for $<\rho^2>,$ which has
a constant value corresponding to the Coulomb phase.
The quantity  $P_s(B)$
varies with the magnetic field, but its variation is trivial, in the
sense that its value is the product of $P_s$ and $\cos B.$ We recall that: 
$$P_{s}(B)=(\cos B)<\cos F_{12}>- (\sin B) <\sin F_{12}>.$$ 
This suggests that the magnetic field penetrates completely and
there is no sign of shielding. The behaviour of the topological number $Q$ is
consistent with this picture: it is zero for all the values of the
magnetic field, indicating that no vortices are formed by the fluctuating
gauge field and the background field penetrates with no obstacles into
the bulk of the lattice.

For $\bt_{hs}=0.500$ the system is in the Higgs phase, as may be seen from
the value of $<\rho^2>.$ For relatively small values of the external field
(up to $n=2$) $P_t$ and $P_s(B),$ as well as $<\rho^2>,$ are approximately
constant, while $P_s$ changes even for small fields. It seems that
the system tunes $F_{12}$ in such a way as to compensate
the change of B, so that $P_{s}(B)$ is constant. Small magnetic fields appear
to get shielded by the dynamics of the scalar and gauge bosons. The
picture changes for bigger external fields: $P_s$ changes and moves towards
a value characterizing the Coulomb phase and this phenomenon is also
confirmed from $<\rho^2>.$ The picture is that too big magnetic fields
cannot be shielded any more and they penetrate into the lattice; then we pass
to the Coulomb phase without Meissner effect.
When we pass to the Coulomb phase $P_s(B)$ satisfies $P_{s}(B)=(\cos B) P_{s}$
approximately. One may check $n=6.$
This picture is further supported by the variation of $Q.$
This number is 1 for $n=1,$, which signals the creation of a vortex
to exactly cancel this magnetic field, while for $n=2$ it equals 2,
shielding the external field completely. For $n=3$ and $n=4$ no new
vortices are created, so the screening is incomplete, and finally,
for even bigger $n,$ even these two vortices disappear.

\[
\begin{array}{||c|c|c|c|c|c|c||}
\hline
\multicolumn{7}{||c||}{{\rm Table~~2}} \\
\hline
\multicolumn{7}{||c||}{{\rm Volume}=8^3,~\z=1.0,~\beta_{gs}=6.0,~x=2.0} \\
\hline
{\rm n} &{\rm cos B} &{\rm P_s} &{\rm P_s (B)} & {\rm P_t} & \rho^2 & Q  \\
\hline
\multicolumn{7}{||c||}{\beta_{hs}=0.200} \\
\hline
 0 &1.0    &0.9429   &0.9248   &0.9429   &1.03 & 0 \\
 4 &0.9238 &0.9428   &0.8711   &0.9429   &1.03 & 0 \\
 6 &0.8314 &0.9427   &0.7838   &0.9429   &1.04 & 0 \\
 8 &0.7071 &0.9429   &0.6667   &0.9429   &1.03 & 0 \\
12 &0.3826 &0.9427   &0.3608   &0.9429   &1.04 & 0 \\
16 &0.00   &0.9427   &0.0000   &0.9429   &1.04 & 0 \\
20 &-0.382 &0.9428  &-0.3608   &0.9429   &1.03 & 0 \\
\hline
\multicolumn{7}{||c||}{\beta_{hs}=0.500} \\
\hline
 0  &1.0     &0.9526   &0.9526   &0.9527  &12.08 & 0 \\
 1  &0.9951  &0.9479   &0.9525   &0.9527  &12.09 & 1 \\
 2  &0.9807  &0.9338   &0.9520   &0.9526  &12.09 & 2 \\
 3  &0.9569  &0.9226   &0.9367   &0.9518  &11.40 & 2 \\
 4  &0.9238  &0.9250   &0.9003   &0.9506  &10.31 & 2 \\
 5  &0.8819  &0.9325   &0.8235   &0.9493  & 9.05 & 0 \\
 6  &0.8314  &0.9325   &0.7765   &0.9488   &8.54 & 0 \\
 8  &0.7071  &0.9332   &0.6612   &0.9480   &7.72 & 0 \\
12  &0.3826  &0.9346   &0.3585   &0.9464   &6.01 & 0 \\
16  &0.00    &0.9376   &0.0004   &0.9455   &4.82 & 0 \\
20 &-0.3826  &0.9394  &-0.3594   &0.9448   &3.90 & 0 \\
\hline
\multicolumn{7}{||c||}{\beta_{hs}=0.700} \\
\hline
 0  &1.0     &0.9570   &0.9570   &0.9571  &13.92 & 0 \\
 2  &0.9807  &0.9380   &0.9563   &0.9569  &13.92 & 2 \\
 4  &0.9238  &0.8825   &0.9550   &0.9568  &13.90 & 4 \\
 6  &0.8314  &0.7920   &0.9522   &0.9562  &13.89 & 6 \\
 7  &0.7730  &0.7349   &0.9502   &0.9559  &13.90 & 7 \\
 8  &0.7071  &0.6706   &0.9475   &0.9555  &13.90 & 8 \\
 9  &0.6343  &0.6904   &0.8905   &0.9546  &13.19 & 7 \\
10  &0.5555  &0.8011   &0.7411   &0.9535  &11.93 & 4 \\
12 & 0.3826  &0.8149   &0.5614   &0.9525  &11.17 & 3 \\
14  &0.1950  &0.8499   &0.2684   &0.9517  &10.36 & 1 \\
16  &0.000   &0.8561   &0.0167   &0.9512  & 9.89 & 0 \\
18 &-0.1950  &0.8593   &-0.1531  &0.9509  & 9.59 & 0 \\
20 &-0.3826  &0.8638   &-0.3175  &0.9507  & 9.41 & 0 \\
\hline
\end{array}
\]

For $\bt_{hs}=0.700$ the system is even deeper than before 
in the Higgs phase, as is obvious from the last column. 
This enables the system to screen the external field
more efficiently: $P_s(B)$ is almost constant up to $n=8,$ as compared 
to $n=2$ previously. The variation of $P_s$ starts early again. The 
penetration of B and the transition to the Coulomb phase is slower 
than before. One may check easily that $P_{s}(B)$ is no longer the product 
of $P_{s}$ and $\cos B$, at least for small $B$; for large $B$, which is the
region where the system approaches the Coulomb phase, the equation 
$P_{s}(B)=(\cos B) P_{s}$ is again approximately satisfied (one may check, 
for example, $n=16$). 
The behaviour of the topological number is interesting here: the system 
creates $Q$ vortices, with $Q=n,$ for $n\le 8.$ As the magnetic 
field increases, the vortex number $Q$ decreases and the shielding 
is incomplete; for $n \ge 16$ no vortex is present any more and the system 
moves towards complete penetration. 

Finally, we note that $P_{t}$ is not very sensitive 
to the presence of the magnetic field anyway.

\[
\begin{array}{|cccccccc|c|cccccccc|}
\hline
\multicolumn{17}{|c|}{Table~3}\\
\hline
\multicolumn{17}{|c|}{q_s(x) {\rm ~at~random~z-coordinates~for~} n=2}\\
\hline
\multicolumn{8}{|c|}{\z=1.0}& & \multicolumn{8}{|c|}{\z=0.1} \\
\hline
0& 0& 0& 0& 0& 0& 0& 0&     & 0& -1& 1& 0& 0& 0& 0& 0\\
0& 0& 0& 0& 0& 0& 0& 0&     &  0& 0& 0& 0& 0& 0& 0& 0\\
0& 0& -1& 0& 0& 0& 0& 0&    &  0& 0& 0& -1& 0& 0& -1& 0\\
0& 0& 0& 0& 0& 0& 0& 0&     &  0& 0& 0& -1& 0& 0& 0& 0\\
0& 0& 0& 0& 0& 0& 0& 0&     & 0& 0& 0& 0& 0& 1& 0& 0\\
0& 0& 0& 0& 0& -1& 0& 0&    & 0& 1& 0& 0& 0& 0& 0& 0\\
0& 0& 0& 0& 0& 0& 0& 0&     &  0& 0& 0& 1& 0& 0& 0& 0\\
0& 0& 0& 0& 0& 0& 0& 0&     &  0& 0& 0& -1& 0& 0& 0& 0\\
\hline
\end{array}
\]

To follow more closely the vortex formation, we have concentrated 
on $n=2$ and studied the topological numbers $q_s(x)$ characterizing 
each plaquette in the $xy$ planes for both the S and the A models for
$\bt_{gs}=6.0,\bt_{hs}=0.700.$ The parameters are chosen such that the system 
is deeply in the Higgs phase; thus, we expect that the system will form 
its own vortices to shield the external magnetic field. These vortices will 
reflect the winding number of the background field, namely $n=2$. 
In table 3 we show these numbers for two $xy$ cross
sections of the lattice (one for $\zeta=1.0$ and one for $\zeta=0.1$); the 
z--coordinates of the planes are chosen at random. It is obvious that 
the system adjusts itself, so that it screens the (small) magnetic 
field. Significantly
more vortices appear in the A model; this is presumably due to energetics, 
since in the A model each layer acts on its own. The most important difference 
is that the (two) vortices appearing in the S model are at the same positions 
for {\em all} the $xy$ planes; on the contrary, the planes in the A model appear 
decoupled and the vortices on each one of them have no relation to the ones 
of its nearest neighbours. In addition, we point out that in the A model the total 
sum of the topological numbers over the cross section need not 
necessarily yield two, since there are big fluctuations; for instance, 
in the case  we show in table 3, it is $+1$. We have also used another definition
\cite{Kajantie} for the winding number, in which one first brings the gauge invariant 
link variables in the interval $(-\pi, +\pi]$ and afterwards follows the same steps 
as above; this definition has the advantage of being {\em additive}, so the
sum over a closed surface, such as the cross section of a lattice should be zero.
The results (not shown) have been that in the S model the two vortices shown 
before are still present, but two more vortices with $Q=-1$ appear, so that the 
sum vanishes; this is also the case for the A model.

\begin{figure}
\centerline{\hbox{\psfig{figure=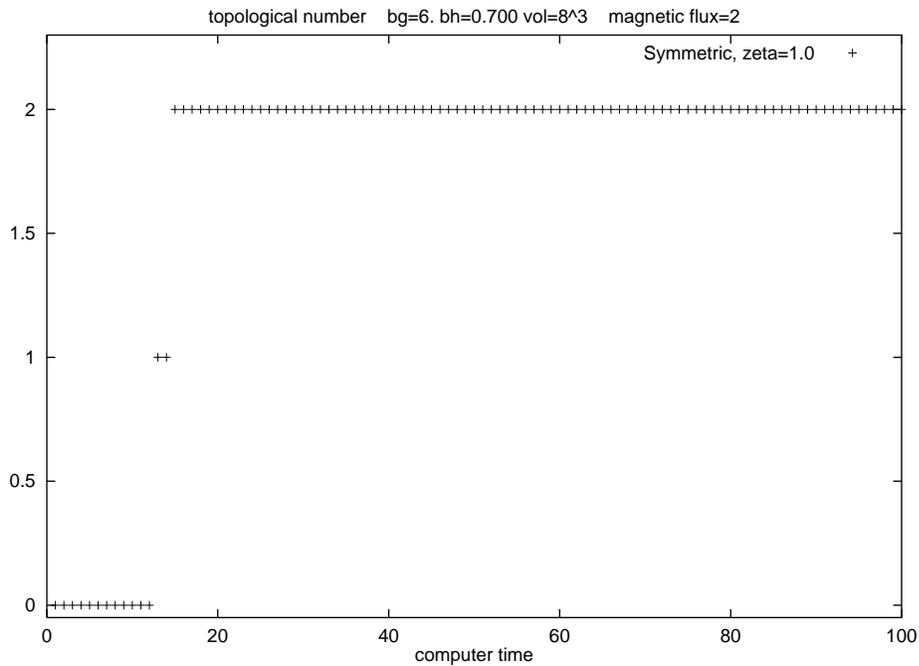,height=9cm,angle=-90}}}
\caption[f8]{Time history of the topological number: $\zeta=1.0, 
~\bt_{gs}=6.0,~\bt_{hs}=0.7,~n=2.$}
\label{f8}
\end{figure}

We illustrate in figure 8 the evolution of the winding
number $Q$ in the S model ($\zeta=1.0$). We see that for the symmetric 
model the topological number changes abruptly and quickly from 0 to 
$n=2$.

The picture is totally different in the A model. In figures 9 and 10 we show
the evolution of Q for two neighbouring layers. We observe that the
topological number changes very slowly and the final value $n=2$ is reached 
only after passing slowly through intermediate values with significant 
fluctuations. Moreover, it is by no means sure that the system has settled in 
its final topological number and that no new fluctuations are going to 
occur. This behaviour is presumably due to the decoupling of the $xy$ 
layers. We have found that the decoupling takes place at $\zeta \simeq 0.6$, 
as one decreases  $\zeta$ from one to zero. 

\begin{figure}
\centerline{\hbox{\psfig{figure=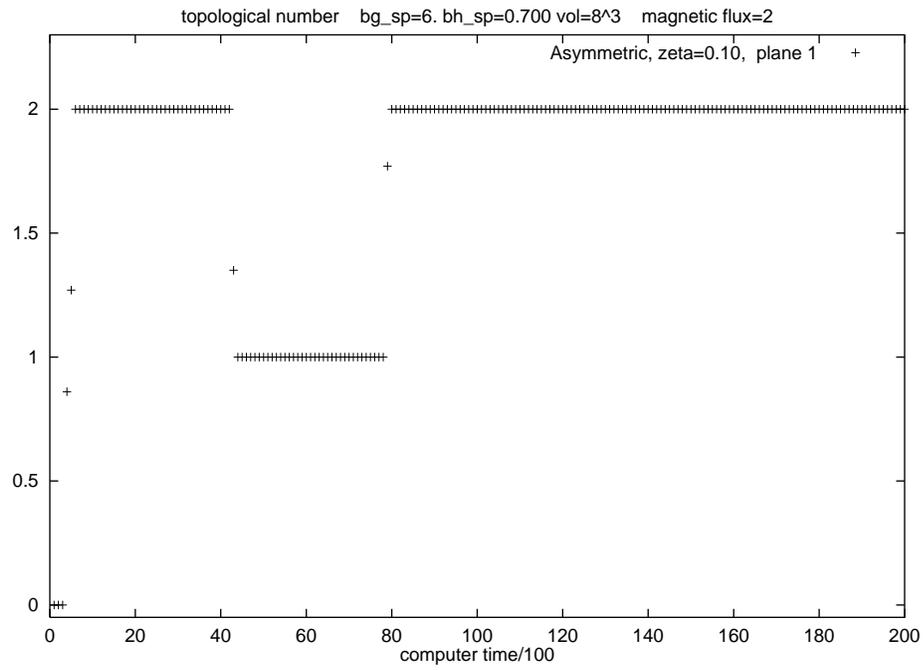,height=9cm,angle=-90}}}
\caption[f9]{Time history of the topological number: $\zeta
=0.1,~\bt_{gs}=6.0,~\bt_{hs}=0.7,~n=2.$}
\label{f9}
\end{figure}

\begin{figure}
\centerline{\hbox{\psfig{figure=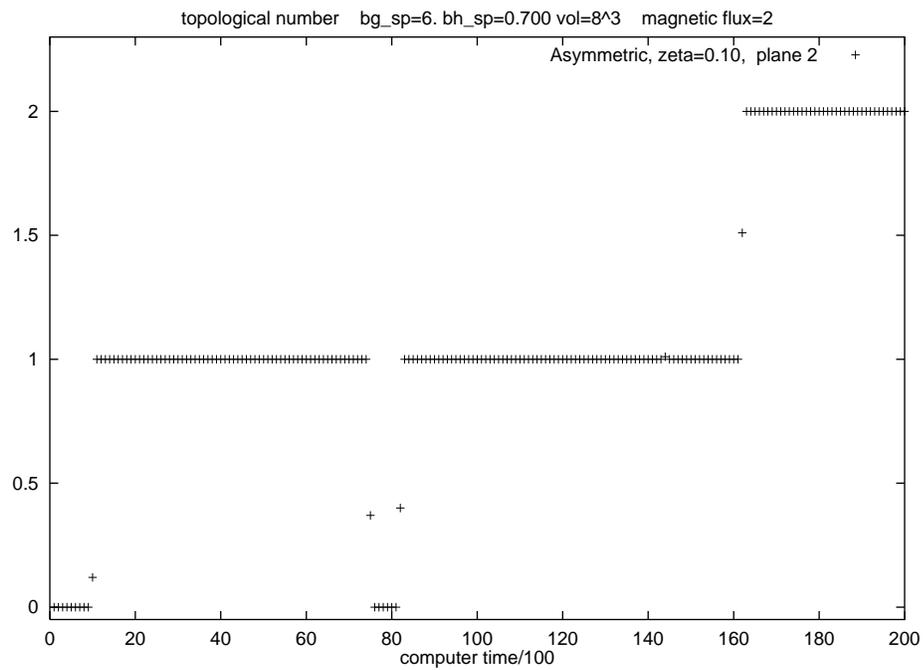,height=9cm,angle=-90}}}
\caption[f10]{Same as in figure 9; neighbouring layer.}
\label{f10}
\end{figure}

In figures 11 and 12 we have shown the evolution of $<\rho^2>$ for the 
S and the A model respectively. They reflect the features of the corresponding 
changes in Q shown in the figures 8, 9 and 10: a sharp and quick increase 
for the S model and a gradual one for the A model, since there the layers 
decouple and do not move together to the new state characterized by 
a non-zero topological number. This may be an indication for the 
generation of the so--called pancake vortices in the layered phase.

\begin{figure}
\centerline{\hbox{\psfig{figure=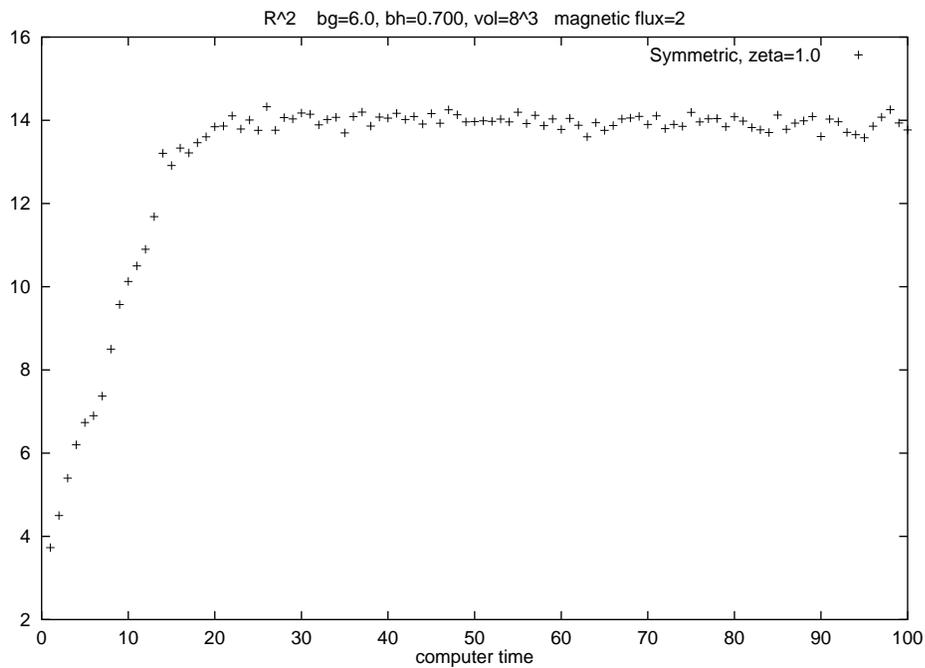,height=9cm,angle=-90}}}
\caption[f11]{Time history of $\rho^2$ for $\zeta=1.0,
~\bt_{gs}=6.0,~\bt_{hs}=0.7,~n=2.$}
\label{f11}
\end{figure}

\begin{figure}
\centerline{\hbox{\psfig{figure=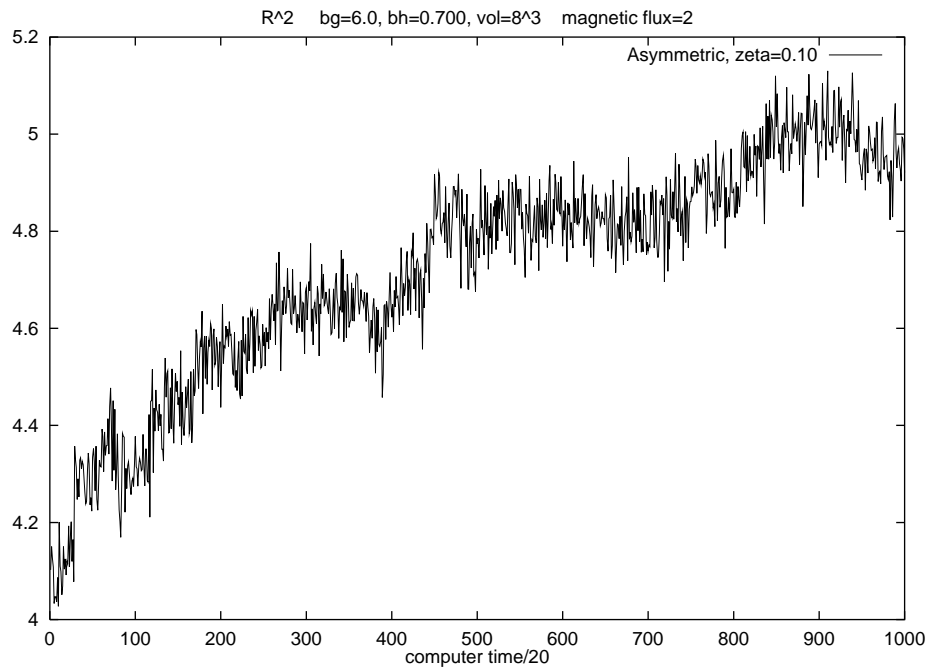,height=9cm,angle=-90}}}
\caption[f12]{Time history of $\rho^2$ for $\zeta=0.1,
~\bt_{gs}=6.0,~\bt_{hs}=0.7,~n=2.$}
\label{f12}
\end{figure}

\vspace{1cm}

{\Large{\bf Acknowledgements}}

\vspace{0.5cm}

The authors would like to thank the TMR project ``Finite temperature
phase transitions in Particle Physics", EU contact number: FMRX-CT97-0122
for financial support. Stimulating discussions with F.Karsch, A. Kehagias, 
C.P. Korthals--Altes, T. Neuhaus, S. Nicolis and N. Tetradis are 
gratefully acknowledged.


\newpage

\begin{thebibliography}{99}
\bibitem {Fu} Y.K.Fu, H.B.Nielsen, Nucl.Phys. B236 (1984) 167;
Nucl.Phys. B254 (1985) 127.
\bibitem {Huang} L.X.Huang, T.L.Chen, Y.K.Fu, Phys.Lett. 329B (1994) 175;
J.Phys.G 21 (1995) 1183; Y.K.fu, L.X.Huang, D.X.Zhang, Phys.Lett. 355B 
(1994) 65.
\bibitem {Berman} D.Berman, E.Rabinovici, Phys.Lett. 157B (1985) 292.
\bibitem {Altes} C.P.Korthals-Altes, S.Nicolis, J.Prades, Phys.Lett. 316B
(1993) 339; A.Huselbos, C.P.korthals-Altes, S.Nicolis, Nucl.Phys. B450 
(1995) 437.
\bibitem{Randall} L.Randall, R.Sundrum, Nucl.Phys. B557 (1999) 79; Phys.Rev.
Lett. 83 (1999) 3370. G.Dvali, M.Shifman, Phys.Lett. B396 (1997) 64, 
Erratum-ibid. B407 (1997) 452, A.Pomarol, hep-ph/9911294 and references 
therein.
\bibitem {Kaplan} D.Kaplan, Phys.Lett. B288 (1992) 342; M.Golterman, K.Jansen,
D.Kaplan, Phys.Lett. B301 (1993) 219; H.Neuberger, R.Narayanan,Phys.Lett.
B302 (1993) 62.
\bibitem{Dorey}H.Kleinert, Gauge fields in Condensed Matter (World 
Scientific 1989); N.Dorey, N.E.Mavromatos, Nucl.Phys. B386 (1992) 614;
K.Farakos, N.E.Mavromatos, Int. J. Mod. Physics B12 (1998) 809.
\bibitem {Dimo} P.Dimopoulos, K.Farakos, G.Koutsoumbas, Eur.Phys.J.C1 (1998)
711.
\bibitem {Kajantie} K.Kajantie, M.Laine, J.Peisa, K.Rummukainen, 
M.Shaposhnikov, Nucl.Phys. B544(1999) 357; K.Kajantie, M.Laine, 
T.Neuhaus, A.Rajantie, K.Rummukainen, Nucl.Phys. B559 (1999) 395;
K.Kajantie, M.Karjalainen, M.Laine, J.Peisa, A.Rajantie, Phys.Lett.B428 
(1998) 334.
\bibitem {Kar} F.Karsch, Nucl.Phys. B205 (1982) 285.
\bibitem {Engels} J.Engels, F.Karsch, H.Satz, I.Montvay, Nucl.Phys. B205
(FS5) (1982) 545.
\bibitem {Montvay} I.Montvay, G.Muenster, Gauge Fields on a Lattice, 
Cambridge University Press, 1994.
\bibitem{Degrand} T.A.DeGrand, D.Toussaint, Phys.Rev.D22 (1980) 2478. 
\bibitem {Dam} P.Damgaard, U.M.Heller, Nucl.Phys. B309 (1998) 625.
\bibitem {Kogut} J.B.Kogut, Rev.Mod.Phys. 51 (1979) 659. 
\bibitem {fkm} K.Farakos, G.Koutsoumbas, N.E.Mavromatos, A.Momen, 
Phys.Rev. D61 (2000) 45005; K.Farakos, G.Koutsoumbas, N.E.Mavromatos, 
Phys.Lett. B431 (1998) 147.
\end{thebibliography}
\end{document}